# Model-based Reliability and Safety: Reducing the complexity of safety analyses using component fault trees

Kai Höfig, Andreas Joanni, Marc Zeller, Francesco Montrone, Martin Rothfelder,
Siemens AG, Corporate Technology, München, Germany.
Rakshith Amarnath, Peter Munk, Arne Nordmann,
Corporate Sector Research and Advance Engineering, Robert Bosch GmbH, Renningen, Germany.

## SUMMARY & CONCLUSIONS

The importance of mission or safety critical software systems in many application domains of embedded systems is continuously growing, and so is the effort and complexity for reliability and safety analysis. Model driven development is currently one of the key approaches to cope with increasing development complexity, in general. Applying similar concepts to reliability, availability, maintainability and safety (RAMS) analysis activities is a promising approach to extend the advantages of model driven development to safety engineering activities aiming at a reduction of development costs, a higher product quality and a shorter time-to-market. Nevertheless, many model-based safety or reliability engineering approaches aim at reducing the analysis complexity but applications or case studies are rare. Therefore we present here a large scale industrial case study which shows the benefits of the application of component fault trees when it comes to complex safety mechanisms. We compare the methodology of component fault trees against classic fault trees and summarize benefits and drawbacks of both modeling methodologies.

## 1 INTRODUCTION

Related work in the area of model-based safety evaluation and reuse of reliability and safety analyses is presented in the next section. We are going to discuss two different systems, that have been renamed since we do not want to allow a link to the real systems but still want to show how real world applications of component fault trees look like. The first system we are going to present is called situation display in section 3 and the second system we are presenting here is called cross link redundancy in section 4. With these two systems, we are able to explain our findings for model-based reliability and safety engineering from the problem domains reuse and complexity. We conclude this paper in section 5.

## 2 RELATED WORK

The use of models in safety engineering processes has gained increasing attention in the last decade and the idea is to support automatic generation of safety artifacts such as fault trees or FMEA tables [2, 3, 4] from a system model. To construct the safety artifact the system models are often annotated with failure propagation models [5, 6]. These failure propagation models are commonly combinatorial in nature thus producing static fault trees. This is also driven by the industrial need to certify [7] their system with static fault trees. Only rarely more advanced safety evaluation models such as Dynamic Fault Trees (DFTs) [8], Generalized Stochastic Petri Nets (GSPNs)[9], State-Event Fault Trees (SEFTs) [10] or Markov models [11]. Beside annotating an architecture specification, there are also approaches to construct a safety artefact via model checking techniques [12].

Nevertheless, large scale examples evaluation the efficiency or the reduction of complexity are rare and publications in this area are reduced to explanatory examples. Therefore in this paper, we summarize our experience for careful reuse of fault tree analysis information encapsulated in component fault trees. It is worth noting that the modeling techniques apply equally to other safety and reliability analysis models.

## 3 CASE STUDY A, THE SITUATION DISPLAY

In this section, we present the first case study of this paper. This case study stems from a real world application and the modeling of the system was done by engineers with some support from modeling experts. Figure 1 shows the elements of the situation display system. Sensor data is an input to the system containing information about the world outside the system. The GPS receiver contains also information from outside the system, but this information comes from a central authority and is not sensed by the situation display itself. This information is transmitted over a redundant channel, collected and compared against each other in the channel interface and then transmitted to the processing component. In processing, sensor data and GPS data are combined to have information of the situation outside the system. This information contains a logical redundancy, since the outside world is evaluated from two different sources, with different information.

Figure 3 shows the classic fault tree analysis of the system for two top events. *Loss of position data* means, that no information about the outside situation is available. This is the case if both signals are lost. *Partial loss of position data* means that either the GPS information is not available or the sensor data is not available. In this case, the analysts decide to make the worst case assumption using an or-gate instead of the, more precise, xor-gate. The top event *erroneous position data* is not depicted but it is shaped like the fault tree for the top event *Loss of position data* with only or-gates and erroneous basic events for each component. It models the situation where the situation data displayed is somehow inaccurate. The analysts decided to make a pessimistic tree for this top event as well where any failure in one of the components can contribute to this top event.

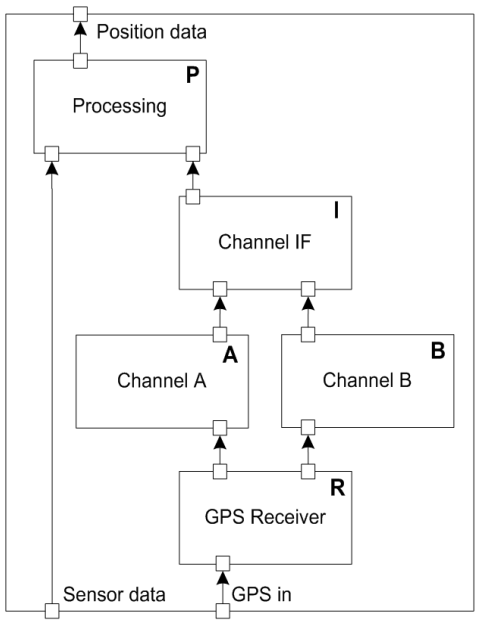

Figure 1: The elements of the situation display system

Figure 2 shows the component fault tree analysis that was made for this system. The component fault tree elements that are related to components of the system are depicted as break out boxes. The CFT for the channel components is redundant and therefore only depicted once. The triangles related at the out port of the processing component model the top events of the system, which are the same as for the classic FTA. *Lo* refers to *loss of position data*, *pLo* refers to *partial loss of position data* and *err* refers to *erroneous position data* accordingly. Component fault trees [13], or more precisely component integrated fault trees [14], facilitate the use of interfaces from the system model. For example, the triangle labeled with *Lo* and *Err* related to the right in port of the component processing are used to model the failure modes *loss of* and *erroneous* contributions that propagate over the interface from component *channel interface* to component *processing*.

During the case study, some findings can be observed. Some of them are related to benefits or drawbacks of the modeling strategies used, other findings can be used as hints for proper modeling.

**Deep trees** It is common practice in a FTA to follow the chain of action backwards through the system and document the intermediate failure modes. An example for this is provided in figure 3, where the fault trees are deep instead of making up a flat structure in a disjunction collecting all combinations that lead to a top event. Since component fault trees follow the system structure we can see here that component fault trees promote such deep structures instead of resulting in flat disjunctions. Component fault trees therefore should start with the top event at the actuator or the component that releases power or where the top event will be visible. The failure analysis is then done backwards or upstream to the participating components until all possible causes for the investigated top-event are identified.

**Localization** When changes have to be made, component fault trees support the search for the right place where changes have to be included in the failure analysis. Changes that can be narrowed to a single component, can also be narrowed to the corresponding component fault tree. In a classic fault tree approach, the right place for changes lies in the hand of the analyst. They can be included in the right place, but the can also be included easily in a disjunctive way by adding additional causes or combination of causes at the uppermost or-gate.

**Redundancies** Redundancies in classic fault trees require the precise identification, which events are related to the redundancy. In a component fault tree approach, redundancies are supported by starting with the doubling of the component fault tree element. Common causes could be problematic and simple doubling should not be applied without a precise analysis for common causes. Nevertheless, in the example mentioned above, the channel components are redundant and the simple doubling of the component fault tree eases the modeling and the analysis of the system.

**Organizational structures** In large scale projects, reliability and safety analysis models are at system level. If the system complexity increases, classic fault trees do not provide a proper divide and conquer strategy to cope with the complexity. In component fault trees, the entire failure behavior (internal) and failure propagation (interfaces) is documented at once. Using such a modeling approach, teams that are organized along the system level components can use this to interact with the other teams not only on the functional interface level of the system model, but also on the interface level for failure propagation. So, component fault trees or similar approaches are a better choice to break down the complexity using the often already existing organizational structures.

**Systematic faults** Since is is important to uncover systematic faults in critical applications, it is also a huge challenge for complex critical systems. Component fault trees using the interfaces from the system model and, as described earlier, if the teams in an organization follow the system decomposition into components, teams can communicate with each other using the interfaces also for failure propagation purposes. This way of diverse communication has potential to uncover systematic faults better than classic monolithic approaches such as classic fault tree analysis. If, for example, the team related to the processing component in the example figures out, that an erroneous signal from the channel interface component can result in an erroneous situation as an output, this team *requests* information about the contribution of the channel interface component from the responsible team. If the team responsible for the channel interface has not thought about the causes for a possible erroneous output, for whatever reason, they have to analyze the component to satisfy the request from the processing component. Vice versa, this scenario goes in both ways. Components can also *provide* information over the interface to other components, that were not investigated so far by the other teams. This back and forth over the interfaces of a system until all provisions and requests are satisfied encourages a more precise discussion about failures and their propagation than a methodology without such failure propagation interfaces.

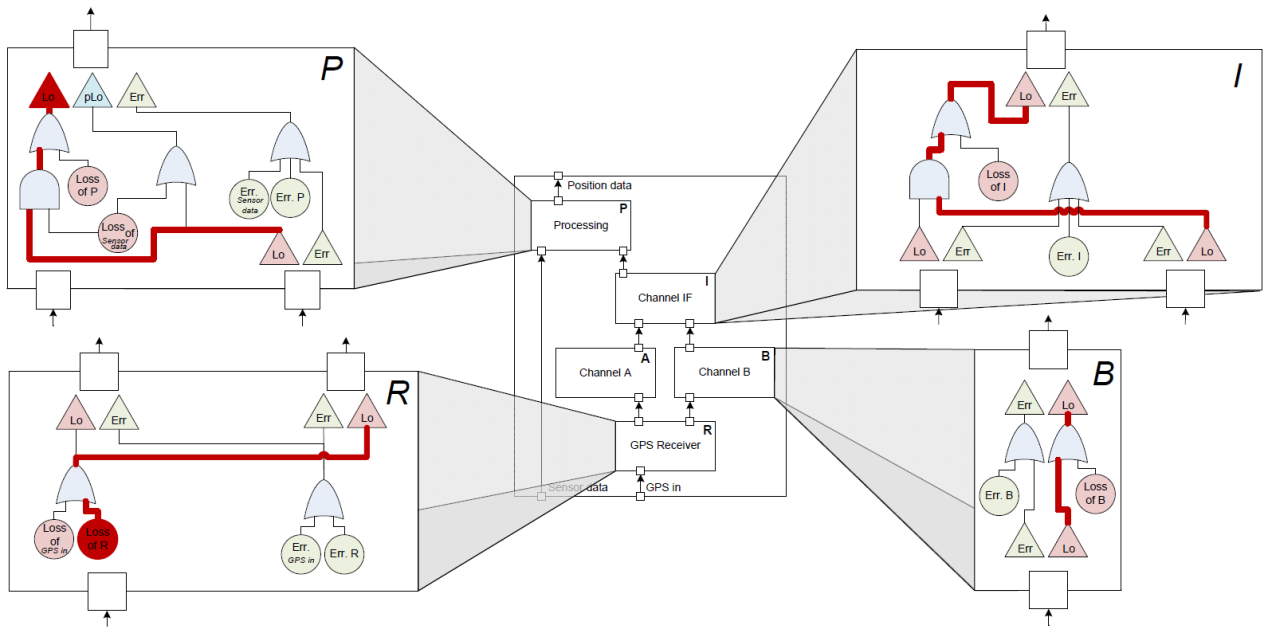

Figure 2: Component fault tree analysis of the situation display system

**Reduction of top events** In the classic fault tree approach as described above, there are three top events. Changing the system requires the identification of all top events that are affected by the change. If the number of top events increases, this can be a time consuming task, which includes the revision of all fault trees. Besides this, there is also a high potential of having the same thoughts more than once, since many fault trees are similar to each other. An example can be found here: the classic fault trees for a *partial loss of* and for a *loss off* look the same, except for one gate. Component fault trees decrease the number of top events on system level at component level to top events that are only relevant for a component. In the example, the top event *partial loss of* only appears at the processing component. If changes are made to the channel interface component, no analysis effort is wasted for the top event *partial loss of*, since is it not relevant for this component. In complex systems with many top events at system level, this benefit can mean a dramatic reduction of complexity.

**Reuse** Reusing a component fault tree element from one system to include it in a new system appears attractive if the component is reused or for components of the shelf. Doing that without any second thoughts about the changed environment and the different context a new system can be very risky. But nevertheless, a reused component fault tree can be a good starting point for an analysis and can help to transfer knowledge about an existing component at the end of a project to the early development phase of a new project. If this information is considered at the beginning of a project, it can also help to prevent systematic faults in the further development activities. If components are used over and over again, the model underlies a maturation process that makes the component fault tree model for this component richer. The potential for uncovering systematic faults with this component is growing accordingly.

## 4 CASE STUDY B, THE CROSS LINK REDUNDANCY

In this section, we present the second case study of this paper. Figure 4 shows the elements of the cross link redundancy. In this two channel system two control units, ECU and CCU, receive steering signals via a wireless connection. The switch component distributes the wireless signals to the ECU and CCU and also distributes the ECU and CCU signals to a pair of actuators, actor 1 and 2. Furthermore, the switch components distribute the ECU and CCU signals crosswise to the other channel via the cross link component. So, if ECU and CCU of channel A and actor 1 and 2 of channel B have a breakdown, the system is still available, since the actors 1 and 2 of channel A are steered with the signals from channel B.

Figure 5 shows the classic fault tree for the system described previously. As can be seen, even to understand the structure of the classic fault tree is complex when it comes to the case where the cross link redundancy is active.

Here, only the component fault tree elements for the components cross link (figure 6) and switch (figure 6) are depicted. All other components are comparatively simple and contain only basic events that directly lead to the top event of the particular component not being available.

The cross link component fault tree is also of comparatively low complexity. It simply propagates the failure mode information about the loss of ECU and CCU from one channel to the other. Additionally, it adds up a failure portion to each failure propagation as a result from internal faults.

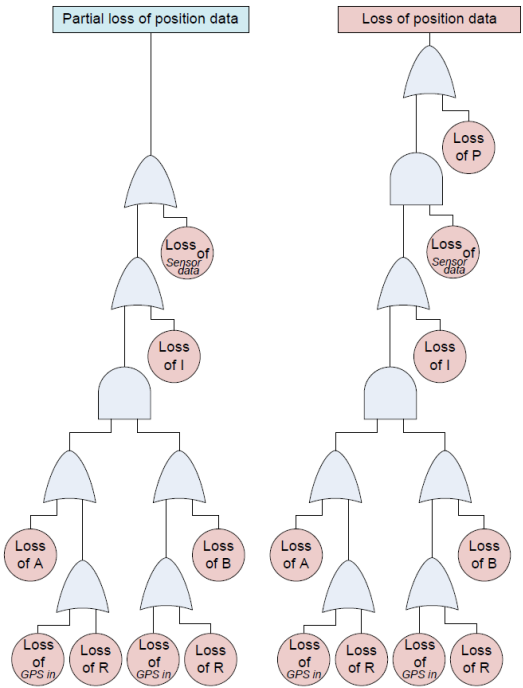

Figure 3: Classic fault tree analysis of the situation display system

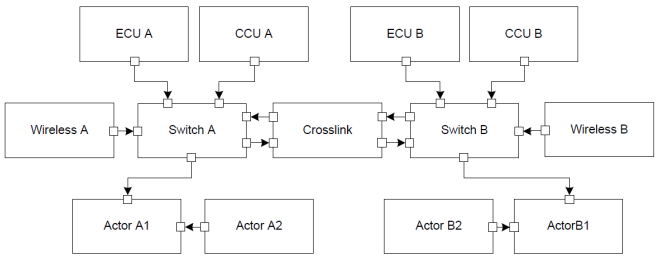

Figure 4: The elements of the cross link redundancy system

A bit more complexity can be found in the component fault tree of the switch. Each switch on every side of the channel propagates the loss of ECU and CCU failure modes of its side to the other side via the cross link component (left side of the figure, black triangles related to the *CL₋ out* port). Additionally it receives failure information from the cross link component about ECU and CCU losses from the other channel from the cross link component (*loss of redundant ECU* and *loss of redundant CCU* related to the port *CL₋ in*). If now either there is no communication input, the switch itself is down or both ECUs and both CCUs are down, the switch reports the failure mode *loss of this channel* to the output of the switch.

The findings described for case study A apply accordingly to this case study. Especially *deep trees*, *localization*, *redundancies*, *organizational structures* and *reuse*. Additionally the following two findings were observed during the second case study.

**Reduction of complexity** The classic fault tree as depicted in figure 5 looks nice and tidy, but it took a while to be sure that this is the tree we want to have. It contains the core idea of the system, but the real world application contained additional top events and in combination with the complex redundancy mechanism, the tree was not easy to handle and the system was hard to analyze. Component fault trees at least did push all the complexity to the switch component and the interfaces did help to bring the thoughts about the redundancy mechanism in the right direction. It was a reduction of complexity and it was also an increase of confidence. In combination with the benefit of having the other components out of the way and contained in separate component fault trees, this was a reduction of complexity by the divide and conquer strategy of component fault trees.

**Need for a holistic model** When applying new modeling strategies such as component fault trees, the established modeling strategies still remain valid, especially in authorities and for approvals to set a system into operation. So the question always is: how can we go back, from the new modeling strategy with all its benefits, to the classic approach required by different stakeholders of a development process? Figure 7 shows the fault tree that is generated from the component fault tree approach for the cross link redundancy system. It was simply generated from following the path through the network of failure propagations from the output of the system to all failure causes. On the one hand, it is logically identical to the fault tree depicted in figure 5, but on the other hand it is not easy to comprehend. Applying a modularized strategy for analysis does not necessarily lead to a better documentation especially from the certification point of view.

## 5 CONCLUSIONS AND FUTURE WORK

In this paper, two case studies of safety and reliability critical systems are presented. Both are analyzed using classic fault trees and component fault trees facilitating interfaces. Findings from the case studies are listed to compare the two different FTA techniques against each other. Since it is hard to prove that one modeling technique provides benefits over another modeling technique, the goal of this paper is to provide references where component fault trees provide potential for the reduction of complexity, increase reuse of modeling artifacts and protection from systematic faults. All findings are described using an example from an industrial application. Since the models presented here are anonymized from real applications, they are called case studies.

For our future work, we aim at more case studies with the goal to get empirical statements to the findings from the case studies.

## REFERENCES

[1] Daniel Ratiu, Marc Zeller, and Lennart Killian. Safety.lab: Model-based domain specific tooling for safety argumentation. *International Workshop on Assurance Cases for Software-Intensive Systems (ASSURE'15)*, 2015.

[2] Tadeusz Cichocki and Janusz Górski. Failure mode and effect analysis for safety-critical systems with software components. In Floor Koornneef and Meine van der Meulen, editors, *Computer Safety, Reliability and Security, 19th International Conference, SAFECOMP*

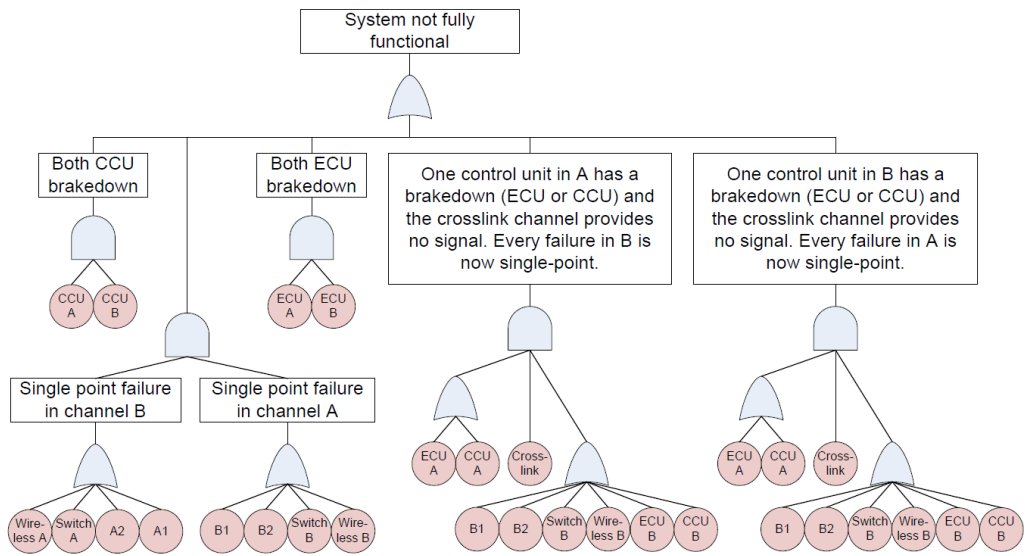

Figure 5: Classic fault tree analysis of the cross link redundancy system

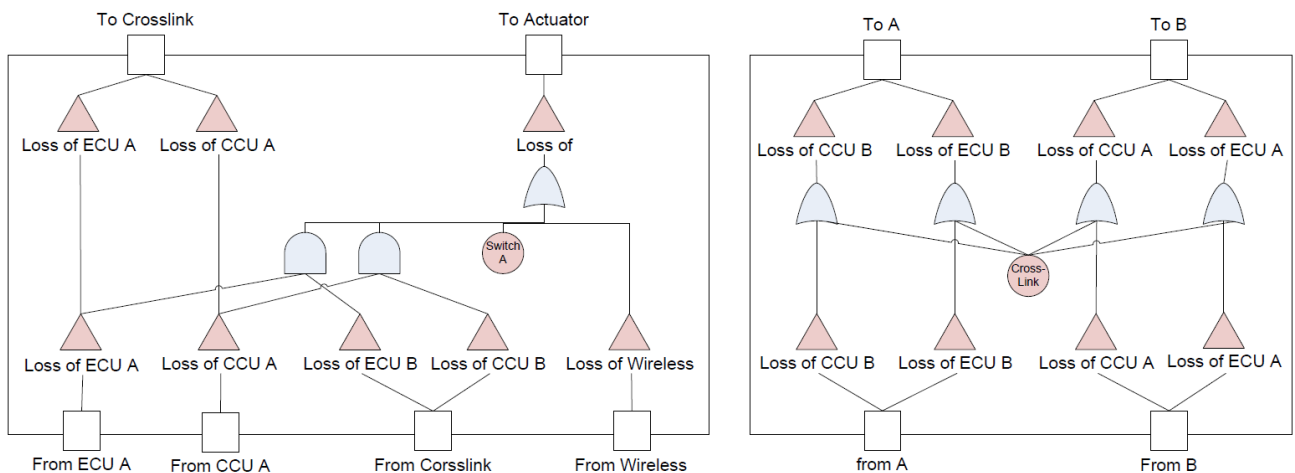

Figure 6: Component fault tree analysis of the switch component of channel A (left) and the crosslink component (right)

2000, Rotterdam, The Netherlands, October 24-27, 2000, Proceedings*, volume 1943 of *Lecture Notes in Computer Science*, pages 382–394. Springer, 2000.

[3] M. Walker, Y. Papadopoulos, D. Parker, and H. Lnn et al. Semi-automatic fmea supporting complex systems with combinations and sequences of failures. *SAE Int. J. Passeng. Cars - Mech. Syst. 2(1)*, pages 791–802, 2009.

[4] Martin Walker and Yiannis Papadopoulos. Qualitative temporal analysis: Towards a full implementation of the fault tree handbook. *Control Engineering Practice*, 17(10):1115 – 1125, 2009.

[5] Dominik Domis and Mario Trapp. Integrating Safety Analyses and Component-Based Design. In *SAFECOMP*, pages 58–71, 2008.

[6] Jonas Elmqvist and Simin Nadjm-Tehrani. Safety-Oriented Design of Component Assemblies using Safety Interfaces. *Formal Aspects of Component Software*, 2006.

[7] IEC61508. International Standard IEC 61508, 1998. International Electrotechnical Commission (IEC).

[8] Joanne Bechta-Dugan, Salvatore Bavuso, and Mark Boyd. Dynamic fault-tree models for fault-tolerant computer systems. *IEEE Transactions on Reliability*, 41(3):363–77, sep 1992.

[9] Ana-Elena Rugina, Karama Kanoun, and Mohamed Kaâniche. A System Dependability Modeling Framework Using AADL and GSPNs. In *Architecting Dependable Systems IV*, volume 4615 of *LNCS*, pages 14–38. Springer, 2007.

[10] Lars Grunske, Bernhard Kaiser, and Yiannis Papadopoulos. Model-driven safety evaluation with state-event-

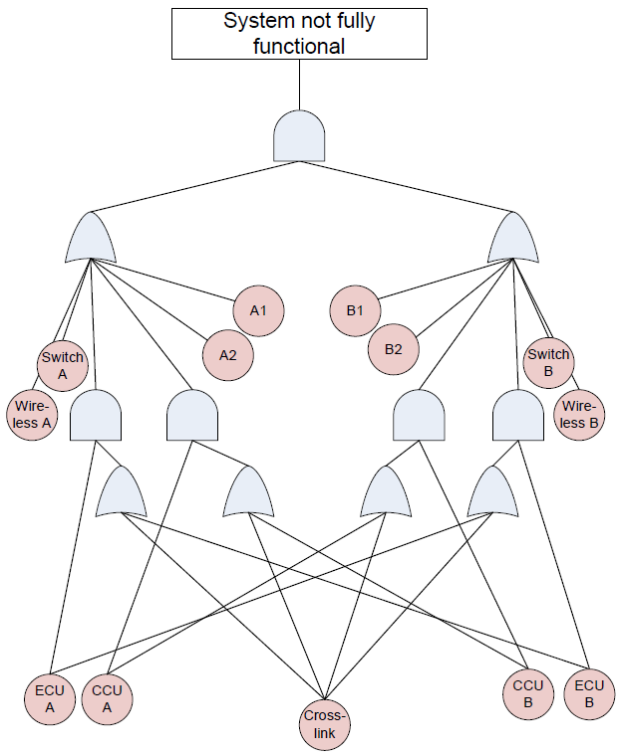

Figure 7: Classic fault tree generated from component fault tree

based component failure annotations. In *8th Int. Symp. on Component-Based Software Engineering, CBSE 2005, Proc.*, pages 33–48, 2005.

[11] Marco Bozzano, Alessandro Cimatti, Joost-Pieter Katoen, Viet Yen Nguyen, Thomas Noll, and Marco Roveri. Safety, dependability and performance analysis of extended aadl models. *Comput. J.*, 54(5):754–775, 2011.

[12] Mats Per Erik Heimdahl, Yunja Choi, and Michael W. Whalen. Deviation analysis: A new use of model checking. *Automated Software Engineering*, 12(3):321–347, 2005.

[13] Bernhard Kaiser, Peter Liggesmeyer, and Oliver Mäckel. A new component concept for fault trees. In *SCS '03: Proceedings of the 8th Australian workshop on Safety critical systems and software*, pages 37–46, Darlinghurst, Australia, 2003. Australian Computer Society, Inc.

[14] Rasmus Adler, Dominik Domis, Kai Höfig, Sören Kemmann, Thomas Kuhn, Jean-Pascal Schwinn, and Mario Trapp. Integration of component fault trees into the uml. In Juergen Dingel and Arnor Solberg, editors, *Models in Software Engineering*, volume 6627 of *Lecture Notes in Computer Science*, pages 312–327. Springer, Berlin / Heidelberg, Germany, 2011.

This project has received funding from the European Unions Horizon 2020 research and innovation programme under grant agreement No 732242, see www.deis-project.eu

BIOGRAPHIES

**Kai Höfig** has a diploma in computer science from RWTH Aachen and a PhD from the University of Kaiserslautern. As a Senior Key Expert he is currently leading the model-based reliability and safety engineering lab at Siemens Corporate Technology.

**Andreas Joanni** is a Senior Key Expert for RAM Analysis and Management at Siemens Corporate Technology and is also active in the IEC TC 56 standardization committee for dependability.

**Marc Zeller** works as a research scientist at Siemens AG, Corporate Technology in the area of model-based safety and reliability engineering. He studied Computer Science at the Karlsruhe Institute of Technology (KIT) and obtained a PhD from the University of Augsburg.

**Francesco Montrone** is a Senior Principle Expert for dependability analysis at Siemens Corporate Technology.

**Martin Rothfelder** has 25+ years experience in safety and reliability. He is heading the Research Group Dependability Analysis and Management of Siemens Corportate Technology.

**Rakshith Amarnath** has a Masters in Embedded Systems from the Delft University of Technology. He is currently working on dependable computing for autonomous driving at Bosch Corporate Research and is also representing Bosch's research topics in the German Federal publicly funded ARAMiS II project.

**Peter Munk** received his PhD (Dr.-Ing.) degree from the Technische Universtitt Berlin in 2016, focusing on software-implemented fault-tolerance mechanisms for real-time applications on multi-core processors. He currently works on model-based safety analysis in the Corporate Sector Research and Advance Engineering of the Robert Bosch GmbH.

**Arne Nordmann** received his PhD (Dr.-Ing.) from Bielefeld University in 2015, focusing on model-driven engineering methods and domain-specific languages in the context of robotics systems. In 2015 he joined the Bosch Corporate Research department to work on model-based safety assessment of highly-automated driving architectures and robotics systems.